\documentclass[twocolumn, pra, superscriptaddress]{revtex4-1}
\usepackage{amssymb}
\usepackage{amsmath}
\usepackage{epsfig}
\usepackage{color}
\usepackage{graphics, graphicx}
\usepackage{bbold}
\usepackage{psfrag}
\usepackage{mathcomp}
\usepackage{subfigure}
\usepackage{verbatim}
\usepackage{color}
\usepackage[colorlinks,citecolor=blue]{hyperref}
\def\cp#1{\mathbf{#1}}

\begin{document}
\title{Quantum fluctuations on top of a $\mathcal{PT}$-symmetric Bose-Einstein Condensate}
\author{Xiaoling Cui}
\email{xlcui@iphy.ac.cn}
\affiliation{Beijing National Laboratory for Condensed Matter Physics, Institute of Physics, Chinese Academy of Sciences, Beijing 100190, China}
\affiliation{Songshan Lake Materials Laboratory , Dongguan, Guangdong 523808, China}
\date{\today}

\begin{abstract}
We investigate the effects of quantum fluctuations in a parity-time($\mathcal{PT}$) symmetric two-species Bose-Einstein Condensate(BEC). It is found  that the $\mathcal{PT}$-symmetry, though preserved by the macroscopic condensate, can be spontaneously broken by its Bogoliubov quasi-particles under quantum fluctuations. The associated $\mathcal{PT}$-breaking transitions in the Bogoliubov spectrum can be conveniently tuned by the interaction anisotropy in spin channels and the strength of $\mathcal{PT}$ potential. In the $\mathcal{PT}$-unbroken regime, the real Bogoliubov modes are generally gapped, in contrast to the gapless phonon mode in Hermitian case. Moreover, the presence of $\mathcal{PT}$ potential is found to enhance the mean-field collapse and thereby intrigue the droplet formation after  incorporating the repulsive force from quantum fluctuations. These remarkable interplay effects of $\mathcal{PT}$-symmetry and interaction can be directly probed in cold atoms experiments, which  shed light on related quantum phenomena in general $\mathcal{PT}$-symmetric systems. 
 \end{abstract}

\maketitle

\section{Introduction} 
The parity-time($\mathcal{PT}$) symmetry governs a fascinating class of non-Hermitian Hamiltonians whose energy spectra can be purely real and bounded below\cite{Bender}, analogous to the Hermitian ones. Nevertheless, very different from the Hermitian counterpart, their eigenstates are generally non-orthogonal and can even coalesce at exceptional points(EPs), where the $\mathcal{PT}$-breaking transition occurs and the spectra after the transition become complex\cite{book}.  The single-particle $\mathcal{PT}$-symmetric Hamiltonian and the associated breaking transitions have been successfully explored earlier in various photonic, electronic and acoustic systems (see reviews\cite{review1, review2}), and recently also in the quantum walk interferometer\cite{QW_PT}, superconducting circuit\cite{SC_PT}, nitrogen-vacancy center\cite{NV_PT}, trapped ions\cite{ion_PT1, ion_PT2} and ultracold gases\cite{Luo, Gadway, Jo}. 

Given the intriguing single-particle property of $\mathcal{PT}$-symmetry, its interplay with interaction  has become a rapidly developing research frontier in recent years\cite{Wunner, Konotop, Mott1, Ueda, Das, Pan, Yu, Zhou, Littlewood, Littlewood2, Mott2, Yi, Zhang, Scheurer}. %For instance, such interplay effect has been %shown to strongly influence the superfluid-Mott transition\cite{..}, 
%utilized to engineer high-order EPs with ultra-sensitivity\cite{Pan},  enhance fermion superfluidity\cite{Zhou, Zhang} and induce intriguing quantum critical phenomena associated with EPs\cite{Ueda, Littlewood,Littlewood2}. 
Previous studies have focused on the interacting $\mathcal{PT}$-symmetric bosons and fermions based on mean-field analyses\cite{Wunner,Konotop, Das,Zhou,Zhang, Pan, Yu}, and the phase transitions and critical phenomena near EPs\cite{Mott1, Ueda, Littlewood, Littlewood2, Mott2}. To date, little has been said about beyond-mean-field effects with $\mathcal{PT}$-symmetry, and whether such effects can generate equally significant quantum phenomena far from EPs. 
%been explored Here we ask the questions, what role is played by the beyond-mean-field effect in $\mathcal{PT}$-symmetric systems, and whether it can generate significant quantum phenomena even far from EP? 
Answering these questions will help to capture the very intrinsic physics in the interplay of interaction and $\mathcal{PT}$-symmetry, which will in turn shed light on related quantum phenomena in a much broader context.

In this work, we explore the effects of quantum fluctuations on top of a $\mathcal{PT}$-symmetric two-species($\uparrow,\downarrow$) Bose-Einstein Condensate(BEC) with the non-Hermitian potential
\begin{equation}
V_{\rm PT}=\Omega (\sigma_x+i\gamma \sigma_z), \label{H_PT}
\end{equation} 
where $\sigma_{\alpha}$($\alpha=x,y,z$) are the Pauli matrices. Obviously $V_{\rm PT}$ commutes with the $\mathcal{PT}$ operator, with $\mathcal{P}$ flipping the spin($\uparrow\leftrightarrow\downarrow$) and $\mathcal{T}$ changing $i$ to $-i$. % the time-reversal operator  invariant under the operation of spin-parity operator   
In the single-particle level, the physics of $V_{\rm PT}$ has been well studied in literature\cite{review1, review2, QW_PT,   SC_PT, NV_PT, ion_PT1, ion_PT2, Luo, Gadway, Jo} and the $\mathcal{PT}$-symmetry is preserved for $\gamma<1$.  
Here we show that when turn on boson-boson interactions,  quantum fluctuations can significantlly affect  the elementary excitation of the system even far from the single-particle EP.  % can lead to intriguing quasi-particle properties that are dramatically different from those in the single-particle or Hermitian cases. 
Specifically, our main findings are listed as below:

{\bf (I)} The $\mathcal{PT}$-symmetry, though preserved by the condensate, can be {\it spontaneously broken} by the Bogoliubov quasi-particles. The $\mathcal{PT}$-breaking transition in the Bogoliubov  spectrum can be conveniently tuned by the strength of $V_{\rm PT}$ and the interaction anisotropy in spin channels. 

{\bf (II)} The quasi-particle in the $\mathcal{PT}$-unbroken regime is generally {\it gapped}, on contrary to the gapless mode in the Hermitian case. Moreover, the mean-field instability of a non-Hermitian system does {\it not necessarily} lead to imaginary excitations therein.  

{\bf (III)} The presence of $V_{\rm PT}$ can enhance the  mean-field collapse of the BEC, and thereby extend the droplet formation to a broader interaction regime than the Hermitian counterpart. % We have identified the parameter regime for such $\mathcal{PT}$-induced droplets to occur and derived their equilibrium densities. 

The experimental relevance of our results and the implication to a general $\mathcal{PT}$-symmetric system  will also be discussed in this paper. 
%These results reveal the fundamental properties of the $\mathcal{PT}$-symmetric BEC, which shed light on more intriguing collective phenomenon due to the interplay of interaction and non-Hermiticity.

The rest of the paper is organized as follows. In section \ref{model} we present the basic model of the system, including the single-particle physics and mean-field treatment. The followed section \ref{mf} is contributed to the mean-field ground state. In section \ref{Bog_analysis}, we build up a systematic theory for the Bogoliubov analysis of the non-Hermitian BEC. The resulted excitation spectrum and the droplet properties are presented, respectively, in section \ref{excitation} and \ref{droplet}. Finally, we discuss the experimental relevance of our results in section \ref{expt} and summarize the whole work in section \ref{summary}.   

\section{Model}  \label{model}

We consider the following Hamiltonian for the interacting two-species bosons under the $\mathcal{PT}$-symmetric potential (we take $\hbar=1$ throughout the paper)
\begin{eqnarray}
H&=&\int d{\cp r} \sum_{\alpha\beta }  \left\{ \Psi^{\dag}_{\alpha}({\cp r})
\big[-\frac{\nabla^2}{2m_{\alpha}}\delta_{\alpha\beta}+\Omega(\sigma^{\alpha\beta}_x+i\gamma\sigma^{\alpha\beta}_z) \big]\Psi_{\beta}({\cp r}) \right. \nonumber\\
&&+ \left. \frac{g_{\alpha\beta}}{2} \Psi^{\dag}_{\alpha}({\cp r})\Psi^{\dag}_{\beta}({\cp r})\Psi_{\beta}({\cp r})\Psi_{\alpha}({\cp r}) \right\}. \label{H}
\end{eqnarray}
Here $\alpha,\beta=\{\uparrow,\downarrow\}$, and $\{\Psi^{\dag}_{\alpha}, \Psi_{\alpha}\}$ are the field operators of spin-$\alpha$ bosons. In order to ensure the $\mathcal{PT}$-symmetry of (\ref{H}), we take the equal mass $m_{\uparrow}=m_{\downarrow}\equiv m$ and equal intra-species coupling $g_{\uparrow\uparrow}=g_{\downarrow\downarrow}\equiv g$. In this case, the property of a homogeneous BEC is determined by three dimensionless parameters:  the dissipation parameter $\gamma$, and two dimensionless combinations $\eta\equiv g_{\uparrow\downarrow}/g$ and $\tilde{\Omega}\equiv \Omega/(gn)$ ($n$ the total density of the BEC).

\subsection{Single-particle physics}

The non-interacting part of (\ref{H}) can be diagonalized as
\begin{equation}
H_0=\sum_{\nu} \epsilon_{\nu; {\cp k}}\Psi^{\dag}_{\nu; {\cp k},R}\Psi_{\nu; {\cp k},L}
\end{equation}
where $\nu=\{+,-\}$ is the index of single-particle eigenstate with eigen-energy  $\epsilon_{\nu; {\cp k}}={\cp k}^2/(2m)+ \nu \Omega\sqrt{1-\gamma^2}$;   $\Psi^{\dag}_{\nu; {\cp k},R}$ ($\Psi_{\nu; {\cp k},L}$) is the associated creation (annilation) operator of the right (left) eigenstate, which satisfies the commutation relation 
\begin{equation}
[\Psi_{\nu'; {\cp k'},L}, \Psi^{\dag}_{\nu; {\cp k},R}]=\delta_{{\cp k}{\cp k}'}\delta_{\nu\nu'}.  \label{relation}
\end{equation}
This relation is equivalent to the bi-orthogonality of right and left eigen-states, which is crucially important for building the theories of non-Hermitian BECs as presented later. 

Since $V_{\rm PT}$ decouples from the kinetic term,  the right/left eigenstates can be decoupled as
\begin{equation}
\Psi^{\dag}_{\nu; {\cp k},R/L}|0\rangle\equiv  |{\cp k}\rangle |\nu\rangle_{R/L},
\end{equation}
where $|0\rangle$ is the vacuum, $|{\cp k}\rangle$ is the plane-wave state with momentum ${\cp k}$, and $|\nu\rangle_{R/L}$ is the spin part of the eigenstate that is solely determined by $V_{\rm PT}$. Specifically, the right and left eigenstates are defined through the Schr{\"o}dinger equations:
\begin{equation}
V_{\rm PT}|\nu\rangle_R=\epsilon_{\nu}|\nu\rangle_R, \ \ \ \ \ \ \ V^{\dag}_{\rm PT}|\nu\rangle_L=\epsilon^*_{\nu}|\nu\rangle_L, \label{schrodinger}
\end{equation}
with $\epsilon_{\nu}=\nu \Omega\sqrt{1-\gamma^2}$. In the regime $\gamma<1$, $|\nu\rangle_{R/L}$ can be expressed as
\begin{eqnarray}
&&|+\rangle_R=C_{+,R}\left(u|\uparrow\rangle + |\downarrow\rangle\right);\ \ \ \ |-\rangle_R=C_{-,R}\left(|\uparrow\rangle -u |\downarrow\rangle \right). \nonumber\\
&& |+\rangle_L=C_{+,L}\left(|\uparrow\rangle +u|\downarrow\rangle\right);\ \ \ \  |-\rangle_L=C_{-,L}\left(u|\uparrow\rangle-|\downarrow\rangle\right),\nonumber\\
\end{eqnarray}
with the parameter
\begin{equation}
u=\sqrt{1-\gamma^2}+i\gamma.
\end{equation} 
Here $C_{\nu,R},C_{\nu,L}$ are all normalization factors. For the Hermitian case ($\gamma=0$ and $u=1$), we can see that the right and left eigenvectors become identical, i.e., $|+\rangle_R\sim |+\rangle_L$, $|-\rangle_R\sim |-\rangle_L$, and different levels are orthogonal to each other $_{R,L}\langle -|+\rangle_{R,L}= 0$. In comparison, for the non-Hermitian case ($\gamma\neq0$ and $u$ is complex), these relations are no longer satisfied, i.e.,  $|\nu\rangle_R\neq |\nu\rangle_L$ and  $_R\langle -|+\rangle_R\neq 0$, $_L\langle -|+\rangle_L\neq 0$. However, given the definition of right/left eigenstates in Eq.\ref{schrodinger}, the bi-orthogonality can be satisfied as long as $\epsilon_+\neq \epsilon_-$:
\begin{equation}
_L\langle -|+\rangle_R=0,\ \ \ \ \ \ \ \ _L\langle +|-\rangle_R= 0. \label{bi_orthogonal}
\end{equation} 
Therefore, the normalization can be carried out between the right and left eigenvectors: 
\begin{equation}
_L\langle \nu|\nu\rangle_R=1,\ \ \ \ \ \ \ \nu=\pm; \label{bi_normalize}
\end{equation} 
which gives 
\begin{equation}
C^*_{\nu,L}C_{\nu,R}=\frac{1}{u+u^*},\ \ \ \ \ \ \ \nu=\pm. \label{CC}
\end{equation}
Note that Eqs.(\ref{bi_orthogonal},\ref{bi_normalize}) guarantees the commutation relation (\ref{relation}).
In this work, we choose a specific gauge such that the normalization factors are all real and identical:
\begin{equation}
C_{\nu,R}=C_{\nu,L}=\frac{1}{\sqrt{u+u^*}}. \label{CC2}
\end{equation}
In this way, when the $\mathcal{PT}$ operator acts on these eigenvectors, we have 
\begin{equation}
\mathcal{PT}|\nu\rangle_R=\nu u^* |\nu\rangle_R;\ \ \ \mathcal{PT}|\nu\rangle_L=\nu u^* |+\rangle_L. \label{PT_eigen}
\end{equation}
This demonstrates that, in the  $\gamma<1$ regime, $|\nu\rangle_{R/L}$ are both the eigenstates of $\mathcal{PT}$-operator with eigenvalue $\nu u^*$. If one chooses a different gauge other than (\ref{CC2}), the eigenvalues in (\ref{PT_eigen}) will be changed. However, we have checked that the gauge choice will not affect the physical quantities studied in this work, given that (\ref{CC}) is satisfied.

\subsection{Mean-field treatment of the $\mathcal{PT}$-symmetric BEC}

In the mean-field framework, we can write down a general coherent ansatz for the right state of the BEC: 
\begin{equation}
|\Psi_0\rangle_R={\cal A} \sum_n \frac{\big(\sum_{\nu}\sqrt{N_{\nu}}e^{i\theta_{\nu}}\Psi^{\dag}_{{\nu}; {\cp k=0},R}\big)^n}{n!}|0 \rangle. \label{right}
\end{equation}
Here  $N_{\nu}$ and $\theta_{\nu}$ are respectively the mean number and the phase of the condensate at level $\nu$. In the regime $\gamma<1$, since the single-particle state $\Psi^{\dag}_{{\nu}; {\cp k},R/L}|0\rangle$ preserves the $\mathcal{PT}$-symmetry, it is natural to require the condensate (\ref{right}) equally preserve such symmetry. Given Eq.(\ref{PT_eigen}), this requirement leads to % is an and $|\Psi_0\rangle_R$ are respectively the eigenstates of $\mathcal{PT}$ with eigen-energies $\nu u^{*}$ and $1$, where $u\equiv\sqrt{1-\gamma^2}+i\gamma$, the requirement then leads to 
\begin{equation}
e^{2i\theta_{\nu}}=\nu u^*. \label{phase}
\end{equation} %Note that although individual $\theta_{\nu}$ is gauge-dependent, the relative phase $\Delta \theta\equiv \theta_+-\theta_-$ is gauge-independent with $e^{2i\Delta \theta}=-1$\cite{supple}.
%\begin{equation}
%e^{2i\Delta \theta}=-1. \label{rel_phase}
%\end{equation}
Following the same strategy, we can obtain the left state of the BEC, $|\Psi_0\rangle_L$, which shares the same form as (\ref{right}) except replacing $\Psi^{\dag}_{{\nu}; {\cp k=0},R}$ by $\Psi^{\dag}_{{\nu}; {\cp k=0},L}$. %The specific gauge choice will not influence the physical quantities studied in this work. 

Given the commutation relation (\ref{relation}) and the coherent ansatz (\ref{right}), we can obtain the following  expectation values under the bi-orthogonal basis:
\begin{widetext}
\begin{eqnarray}
&&_L\langle\Psi_0|\Psi^{\dag}_{{\nu}; {\cp k=0},R}|\Psi_0\rangle_R= \sqrt{N_{\nu}}e^{-i\theta_{\nu}},\ \ \ _L\langle\Psi_0|\Psi_{{\nu}; {\cp k=0},L}|\Psi_0\rangle_R= \sqrt{N_{\nu}}e^{i\theta_{\nu}};\nonumber\\
&&_L\langle\Psi_0|\Psi^{\dag}_{{\nu}; {\cp k=0},R} \Psi_{{\nu'}; {\cp k=0},L}|\Psi_0\rangle_R= \sqrt{N_{\nu}N_{\nu'}}e^{i(\theta_{\nu'}-\theta_{\nu})};\nonumber\\
&&_L\langle\Psi_0|\Psi^{\dag}_{{\nu}; {\cp k=0},R}\Psi^{\dag}_{{\nu'}; {\cp k=0},R} \Psi_{{\nu''}; {\cp k=0},L}\Psi_{{\nu'''}; {\cp k=0},L}|\Psi_0\rangle_R=\sqrt{N_{\nu}N_{\nu'}N_{\nu''}N_{\nu'''}}e^{i(\theta_{\nu'''}+\theta_{\nu''}-\theta_{\nu'}-\theta_{\nu})}. \label{expectation}
\end{eqnarray} 
\end{widetext}
This shows that in the mean-field framework under the bi-orthogonal basis, one can replace the zero-momentum operators $\Psi^{\dag}_{{\nu}; {\cp k=0},R}$ and $\Psi_{{\nu}; {\cp k=0},L}$ with their mean values:
\begin{equation}
\Psi^{\dag}_{{\nu}; {\cp k=0},R}\rightarrow \sqrt{N_{\nu}}e^{-i\theta_{\nu}},\ \ \ \Psi_{{\nu}; {\cp k=0},L}\rightarrow \sqrt{N_{\nu}}e^{i\theta_{\nu}}. \label{eq_mf}
\end{equation} 
In this way, we can go on to study the mean-field ground state and examine the effects of quantum fluctuations on top of it. 

Here we would like to emphasize that %it is necessary to use the bi-orthogonal basis for 
the mean-field treatment is only valid under the bi-orthogonal basis, but not if only use one of the basis (right or left).  For instance, we cannot obtain the expectation values as the form in Eq.(\ref{expectation}) if only under the right basis ($_R\langle...\rangle_R$) or the left basis ($_L\langle...\rangle_L$), and as a result we cannot replace the operators by their according mean-field values as in (\ref{eq_mf}). 

\subsection{Interaction channels}

To facilitate later discussions, % on mean-field ground state and quantum fluctuations on top of it
we rewrite the interaction part of (\ref{H}) in the following form:
\begin{equation}
U=\sum_{\nu_1 \nu_2\nu_3 \nu_4} U_{\nu_1 \nu_2;\nu_3 \nu_4} \sum_{{\cp Q} {\cp k} {\cp k'}}\Psi^{\dag}_{\nu_1; {\cp Q}-{\cp k},R}\Psi^{\dag}_{\nu_2; {\cp k},R}\Psi_{\nu_3; {\cp k'},L}\Psi_{\nu_4; {\cp Q}-{\cp k'},L}.
\end{equation}
Here $U_{\nu_1 \nu_2;\nu_3 \nu_4}$ is invariant under the permutation of $\nu_1\leftrightarrow\nu_2$ or $\nu_3\leftrightarrow\nu_4$, and thus there are totally nine different coupling channels,  with five even-parity combinations $\{\nu_1 \nu_2;\nu_3 \nu_4\}=\{++;++\}, \ \{--;--\},\ \{++;--\},\ \{--;++\},\ \{+-;+-\}$, and four odd-parity ones $\{+-;++\}, \ \{+-;--\}, \ \{++;+-\}, \ \{--;+-\}$. The coupling constants in these channels are:  
\begin{eqnarray}
&&U_{++;++}=U_{--;--}=\frac{g}{V}\frac{u_-}{4};\nonumber\\
&&U_{++;--}=U_{--;++}=\frac{g}{V} u'_+;\nonumber\\
&&U_{+-;+-}=\frac{g}{V} u_+;  \nonumber\\
&&U_{+-;++}=-U_{+-;--}=U_{++;+-}=-U_{--;+-}=-\frac{g}{V} u'',\nonumber\\ \label{U}
\end{eqnarray} 
with 
\begin{eqnarray}
&&u_-=4u'_-=\frac{1-2\gamma^2+\eta}{1-\gamma^2};\ \ \ \ \ \ u_+=\frac{1-\eta\gamma^2}{1-\gamma^2};\nonumber\\
&&u'_+=\frac{1-\eta}{4(1-\gamma^2)}; \ \ \ \ \ \ \ \ \ \ u''= -\frac{i\gamma(1-\eta)}{2(1-\gamma^2)}.
\end{eqnarray} 
Here $u_-,\ u_+$ and $u_+'$ are the coupling constants for even-parity channels, and $u''$ represents the coupling for odd-parity ones. $u''$ is non-zero and purely imaginary only for the non-Hermitian case with spin-independent interaction, i.e., when $\gamma\neq 0$ and $\eta\neq 1$. 
As shown later, the presence of these odd-parity channels will greatly affect the elementary excitation of the BEC.

\bigskip

\section{Mean-field ground state} \label{mf}

To determine the mean-field ground state, we examine the total mean-field energy $E_{\rm mf}=_L\langle \Psi_0|H|\Psi_0\rangle_R$. It is found that under the phase constraint (\ref{phase}),  $E_{\rm mf}$ solely depends on the parameter $x\equiv N_-/N$, where $N=N_-+N_-$ is the total number. Explicitly, the energy per particle $\epsilon_{\rm mf}\equiv E_{\rm mf}/N$ reads
\begin{eqnarray}
\epsilon_{\rm mf}(x)&=&\Omega\sqrt{1-\gamma^2} (1-2x) + \nonumber \\
&&\frac{gn}{1-\gamma^2} \left( \gamma^2(\eta-1)(x^2-x) +\frac{1-2\gamma^2+\eta}{4}\right).
\end{eqnarray}
%It turns out that $\epsilon_{\rm mf}$ is independent of the averaged phase $\theta$. %In the regime $\gamma^2(\eta-1)<2\tilde{\Omega}(1-\gamma^2)^{3/2}$, 
For simplicity, in this work we will focus on the $\eta<1$ regime, where the minimum of $\epsilon_{\rm mf}(x)$ locates at $x=1$, i.e., the bosons condense at the lower branch with energy 
\begin{eqnarray}
\epsilon_{\rm mf}=-\Omega\sqrt{1-\gamma^2} +\frac{gn}{1-\gamma^2} \frac{1-2\gamma^2+\eta}{4}.\label{e_mf}
\end{eqnarray}
%which  is composed by the single-particle energy and the interaction energy of BEC solely in the lower branch. 
Accordingly, we can obtain the chemical potential $\mu\equiv \partial E_{\rm mf}/\partial N$ and further the %, reads 
%\begin{eqnarray}
%\mu=-\Omega\sqrt{1-\gamma^2} +\frac{gn}{1-\gamma^2} \frac{1-2\gamma^2+\eta}{2}. \label{mu}
%\end{eqnarray}
compressibility $\chi\equiv \partial n/\partial \mu$ as
\begin{equation}
\chi=\frac{2}{g}\frac{1-\gamma^2}{1-2\gamma^2+\eta}. \label{chi}
\end{equation}
The mean-field stability against density fluctuations would require $\chi>0$ and therefore
 \begin{equation}
\eta>2\gamma^2-1. \label{stable}
\end{equation}
This condition is more stringent than the Hermitian case ($\eta>-1$).  
In other words, a non-Hermitian BEC (with finite $\gamma$) can undergo mean-field collapse more easily than its Hermitian counterpart ($\gamma=0$). This will be responsible for the $\gamma$-induced droplet formation as discussed later.

\section{Bogoliubov analysis} \label{Bog_analysis}

Given the $\mathcal{PT}$-symmetric BEC at ${\cp k}=0$ and $\nu=-$, we now study its elementary excitations due to quantum fluctuations. Following the standard Bogoliubov approach, we assume $\Psi^{\dag}_{\nu; {\cp k}, R}$ and $\Psi_{\nu; {\cp k},L}$ (except for $\{\nu=-,\ {\cp k}=0\}$) are all small fluctuation operators and only keep in the Hamiltonian all the bi-linear terms of these operators, which gives $H=N\epsilon_{\rm mf}+H_{\rm BG}$ with
\begin{widetext}
\begin{eqnarray}
H_{\rm BG}&=&\sum_{\cp k}\sum_{\nu}\left(  (\epsilon_{\nu; {\cp k}}-\mu+gn u_{\nu})\Psi^{\dag}_{\nu; {\cp k}, R}\Psi_{\nu; {\cp k}, L} +gn u'_{\nu} (e^{2i\theta_-} \Psi^{\dag}_{\nu; {\cp k}, R}\Psi^{\dag}_{\nu; -{\cp k}, R} + e^{-2i\theta_-} \Psi_{\nu; {\cp k}, L}\Psi_{\nu;-{\cp k}, L}) \right) \nonumber\\
&&\ \ \ \  +gn u'' \sum_{\cp k}  \left( e^{2i\theta_-} \Psi^{\dag}_{+; {\cp k}, R}\Psi^{\dag}_{-; -{\cp k}, R} +  e^{-2i\theta_-} \Psi_{+; {\cp k}, L}\Psi_{-; -{\cp k}, L}  +2 \Psi^{\dag}_{+; {\cp k}, R}\Psi_{-; {\cp k}, L} +  2\Psi^{\dag}_{-; {\cp k}, R}\Psi_{+; {\cp k}, L} \right).  \label{Bog}
\end{eqnarray}
\end{widetext}
Here $H_{\rm BG}$ naturally inherits $\mathcal{PT}$-symmetry from the full Hamiltonian (\ref{H}), since we have taken the condensate (\ref{right}) as $\mathcal{PT}$-symmetric. The first line in $H_{\rm BG}$ is reduced from even-parity channels, and the second line from odd-parity ones. Obviously, the effect of odd-parity channels is to couple fluctuations in different branches ($-\leftrightarrow +$), and the coupling constant $u''$ is purely imaginary in the presence of both non-Hermiticity and interaction anisotropy. 
%Obviously, the effect of odd-parity channels is to couple the fluctuations from different branches in  odd-parity channels  We can see that in the case of $u''=0$, i.e., there is no asymmetric channels such as in the Hermitian case ($\gamma=0$) or spin-independent interaction($\eta=1$), the fluctuations of $-$- and $+$-branches are well decoupled. However, in general cases, the fluctuations in two branches are  coupled with each other, and the eigen-Bogoliubov modes can be found as follows. 

To facilitate the diagonalization of the bilinear Hamiltonian (\ref{Bog}), we rewrite it as 
\begin{align}
&H_{\rm BG}=\sum'_{\cp k} \left\{ F_{\cp k}^T {\cal M}({\cp k})G_{\cp k} - \sum_{\nu} (\epsilon_{\nu; {\cp k}}-\mu+gn u_{\nu})\right\},  \label{h_bg1}
\end{align}
where $\sum'$ implies the summation be taken over half of ${\cp k}$-space to avoid the double counting; the vectors are
\begin{align}
&F_{\cp k}=\left(\begin{array}{c}\Psi^{\dag}_{-; {\cp k}, R} \\ \Psi_{-; -{\cp k}, L} \\ \Psi^{\dag}_{+; {\cp k}, R} \\ \Psi_{+; -{\cp k}, L}\end{array}\right),\ \ \ \ G_{\cp k}=\left(\begin{array}{c}\Psi_{-; {\cp k}, L} \\\Psi^{\dag}_{-; -{\cp k}, R} \\ \Psi_{+; {\cp k}, L} \\ \Psi^{\dag}_{+; -{\cp k}, R}\end{array}\right);
\end{align}
and the matrix ${\cal M}$ is
\begin{widetext}
\begin{align}
&{\cal M}({\cp k})=\left(\begin{array}{cccc}\epsilon_{-; {\cp k}}-\mu+gn u_{-}  & 2gnu'_-e^{2i\theta_-}  & 2gnu''   &  gnu'' e^{2i\theta_-} 
\\ 2gnu'_-e^{-2i\theta_-}  & \epsilon_{-; {\cp k}}-\mu+gn u_{-}  & gnu'' e^{-2i\theta_-}   &   2gnu''
\\ 2gnu'' & gnu'' e^{2i\theta_-}    & \epsilon_{+; {\cp k}}-\mu+gn u_{+}  &   2gnu'_+e^{2i\theta_-}  
\\ gnu'' e^{-2i\theta_-}  & 2gnu''  & 2gnu'_+ e^{-2i\theta_-}    & \epsilon_{+; {\cp k}}-\mu+gn u_{+} \end{array}\right). 
\end{align}
\end{widetext}

We aim to diagonalize $H_{\rm BG}$ as the following form:
\begin{align}
&H_{\rm BG}=\sum'_{\cp k}  \tilde{F}_{\cp k}^T \left(\begin{array}{cccc} E_{1{\cp k}} &   &   &   \\  &  E_{2{\cp k}} &   &   \\  &   & E_{3{\cp k}}  &   \\  &   &   & E_{4{\cp k}} \end{array}\right)
\tilde{G}_{\cp k} + {\rm const},  \label{h_bg2}
\end{align}
where $E_{i{\cp k}}$ are the four eigen-modes for Bogoliubov quasi-particles, and the two eigen-vectors are
\begin{align}
&\tilde{F}_{\cp k}=\left(\begin{array}{c}\alpha^{\dag}_{1, {\cp k}, R} \\ \alpha_{2,{\cp k}, L} \\ \alpha^{\dag}_{3, {\cp k}, R} \\ \alpha_{4, {\cp k}, L}\end{array}\right),\ \ \ \ \tilde{G}_{\cp k}=\left(\begin{array}{c}\alpha_{1, {\cp k}, L} \\ \alpha^{\dag}_{2,{\cp k}, R} \\ \alpha_{3, {\cp k}, L} \\ \alpha^{\dag}_{4,{\cp k}, R}\end{array}\right).
\end{align}
The eigen-operators are required to satisfy the commutation relation
\begin{equation}
[\alpha_{i, {\cp k}, L},\alpha^{\dag}_{j, {\cp k}', R}]=\delta_{ij}\delta_{{\cp k}{\cp k}'},\ \ \ \ \ i,j=1,2,3,4. \label{relation2}
\end{equation}

To find out eigen-spectra $E_{i{\cp k}}$ as well as the relation between $\tilde{F}_{\cp k},\tilde{G}_{\cp k}$ and $F_{\cp k},G_{\cp k}$, we start from the equation of motions(EoM) of these vectors.  Based on the Heisenberg equation for non-Hermitian system (see derivation in Appendix \ref{Heisenberg}), we can write down the EoM of $G_{\cp k}$ and $\tilde{G}_{\cp k}$: 
\begin{align}
& i\frac{\partial}{\partial t} G_{\cp k} = \left(\begin{array}{cccc} 1 &   &   &   \\  & -1  &   &   \\  &   &  1 &   \\  &   &   & -1 \end{array}\right) {\cal M}({\cp k}) G_{\cp k};\nonumber\\
&i\frac{\partial}{\partial t} \tilde{G}_{\cp k} = \left(\begin{array}{cccc} E_{1{\cp k}} &   &   &   \\  & -E_{2{\cp k}}  &   &   \\  &   &  E_{3{\cp k}} &   \\  &   &   & -E_{4{\cp k}} \end{array}\right) \tilde{G}_{\cp k}.   \label{eom1} 
\end{align}
This implies that by diagonalizing the matrix ${\rm Diag}(1,-1,1,-1){\cal M}({\cp k})$, we can obtain the four Bogoliubov modes from its eigen-energies. % as $E_{1{\cp k}}$, $-E_{2{\cp k}}$, $E_{3{\cp k}}$ and $-E_{4{\cp k}}$. 
Explicitly, by introducing a transformation matrix ${\cal A}$ in $G_{\cp k}={\cal A}\tilde{G}_{\cp k}$, we have
\begin{align}
& \ \ \ \ \ {\cal A}^{-1} \left[ \left(\begin{array}{cccc} 1 &   &   &   \\  & -1  &   &   \\  &   &  1 &   \\  &   &   & -1 \end{array}\right) {\cal M}({\cp k}) \right] {\cal A} \nonumber\\
&=  \left(\begin{array}{cccc} E_{1{\cp k}} &   &   &   \\  & -E_{2{\cp k}}  &   &   \\  &   &  E_{3{\cp k}} &   \\  &   &   & -E_{4{\cp k}} \end{array}\right). \label{diag1}
\end{align}

Similarly, we can write down the EoM for $F_{\cp k}$ and $\tilde{F}_{\cp k}$, and by introducing a transformation matrix ${\cal B}$ in $F^T_{\cp k}=\tilde{F}^T_{\cp k}{\cal B}$, we have
\begin{align}
& \ \ \ \ \ {\cal B} \left[  {\cal M}({\cp k}) \left(\begin{array}{cccc} -1 &   &   &   \\  & 1  &   &   \\  &   &  -1 &   \\  &   &   & 1 \end{array}\right) \right] {\cal B}^{-1} \nonumber\\
&=  \left(\begin{array}{cccc} -E_{1{\cp k}} &   &   &   \\  & E_{2{\cp k}}  &   &   \\  &   &  -E_{3{\cp k}} &   \\  &   &   & E_{4{\cp k}} \end{array}\right) \label{diag2}
\end{align}
Therefore, the Bogoliubov modes can also be obtained by diagonalizing the matrix ${\cal M}({\cp k}){\rm Diag}(-1,1,-1,1)$. 

The two diagonalization schemes, i.e., one is based on (\ref{diag1}) and the other is based on (\ref{diag2}), produce the same solution of $E_{i{\cp k}}$, which satisfy
\begin{equation}
E_{\cp k}= \sqrt{\frac{-b_{\cp k}\pm \sqrt{b_{\cp k}^2-4c_{\cp k}}}{2}}, \label{E_k}
\end{equation}
with 
\begin{widetext}
\begin{align}
&b_{\cp k}=-(\epsilon_{-; {\cp k}}-\mu+gn u_{-} )^2-(\epsilon_{+; {\cp k}}-\mu+gn u_{+})^2 -(gn)^2\left(6u''^2 - 4u'^2_+ - 4u'^2_- \right); \nonumber\\
&c_{\cp k}=\left[ (\epsilon_{-; {\cp k}}-\mu+gn u_{-} )^2- (2gnu'_-)^2 \right]\left[(\epsilon_{+; {\cp k}}-\mu+gn u_{+} )^2- (2gnu'_+)^2 \right]+(gnu'')^2\left[9(gnu'')^2-40(gn)^2u'_-u'_+ \right. \nonumber\\
& \ \ \ \ \ \ \left.-10(\epsilon_{-; {\cp k}}-\mu+gn u_{-} ) (\epsilon_{+; {\cp k}}-\mu+gn u_{+} )+16gnu'_+(\epsilon_{-; {\cp k}}-\mu+gn u_{-} )+ 16gnu'_-(\epsilon_{+; {\cp k}}-\mu+gn u_{+} )  \right].
\end{align}
\end{widetext}
The four eigen-modes in (\ref{E_k}) fall into two identical pairs, and we choose $E_{1{\cp k}}=E_{2{\cp k}}$ and $E_{3{\cp k}}=E_{4{\cp k}}$. This is also a natural choice since in non-interacting limit, ${\cal M}({\cp k})$ can exactly reduce to the diagonal matrix ${\rm Diag}(E_{1{\cp k}},E_{2{\cp k}}, E_{3{\cp k}},E_{4{\cp k}})$ with $E_{1{\cp k}}=E_{2{\cp k}}$ and $E_{3{\cp k}}=E_{4{\cp k}}$. 

In fact, based on the commutation relations (\ref{relation}) and (\ref{relation2}), we can find out the relation between the two transformation matrixes:
\begin{align}
{\cal A} \left(\begin{array}{cccc} 1 &   &   &   \\  & -1  &   &   \\  &   &  1 &   \\  &   &   & -1 \end{array}\right) {\cal B} = \left(\begin{array}{cccc} 1 &   &   &   \\  & -1  &   &   \\  &   &  1 &   \\  &   &   & -1 \end{array}\right),
\end{align} 
and then one can prove straightforwardly that ${\cal B} {\cal M}({\cp k}) {\cal A}={\rm Diag}(E_{1{\cp k}},E_{2{\cp k}}, E_{3{\cp k}},E_{4{\cp k}})$. It follows that the first term in Eq.(\ref{h_bg1}) is equal to the first term in Eq.(\ref{h_bg2}). Therefore the constant terms in (\ref{h_bg1}) and (\ref{h_bg2}) are also identical. Now we can rewrite Eq.(\ref{h_bg2}) as
\begin{eqnarray}
H_{\rm BG}&=&\sum'_{{\cp k}} \sum_{i=1}^4 E_{i {\cp k}} \alpha^{\dag}_{i, {\cp k},R}\alpha_{i, {\cp k},L} + \nonumber\\
&& \frac{1}{2}\sum_{\cp k} \Big\{ E_{1 {\cp k}} + E_{3 {\cp k}} - \sum_{\nu} (\epsilon_{\nu; {\cp k}}-\mu+gn u_{\nu})\Big\}, \label{diag}
\end{eqnarray}
Further incorporating the regularization of bare couplings $g$ and $g_{12}$ from the mean-field interaction energy (\ref{e_mf}), we can obtain the Lee-Huang-Yang(LHY) energy as:
\begin{eqnarray}
E_{\rm LHY}&=&\frac{1}{2}\sum_{\cp k} \Big\{ E_{1 {\cp k}} + E_{3 {\cp k}} - \sum_{\nu} (\epsilon_{\nu; {\cp k}}-\mu+gn u_{\nu}) \nonumber\\
&&\ \ \ \ +  \frac{1-2\gamma^2+\eta^2}{2(1-\gamma^2)}\frac{m(gn)^2}{{\cp k}^2}\Big\}. \label{E_LHY}
\end{eqnarray}
We have checked that the summation in above equation converges at large ${\cp k}$ and the ultraviolet divergence can be avoided. 

\section{Excitation spectrum}  \label{excitation}

In this section, we present the result of Bogoliubov excitation spectrum for the $\mathcal{PT}$-symmetric BEC. Since $E_{1{\cp k}}=E_{2{\cp k}}$ and $E_{3{\cp k}}=E_{4{\cp k}}$, we will only show the results of $E_{1{\cp k}}$ and $E_{3{\cp k}}$.

To highlight the effect of non-Hermiticity to Bogoliubov excitations, we first go through the Hermitian case ($\gamma=0$). In this case, all odd-parity terms in (\ref{Bog}) are absent ($u''=0$) and the fluctuations in $+$ and $-$ branches are well decoupled. This leads to a gapless spectrum $E_{1{\cp k}}=\sqrt{({\cp k}^2/2m)^2+2\mu_-{\cp k}^2/(2m)}$ and a gapped one $E_{3{\cp k}}=\sqrt{({\cp k}^2/2m+2\Omega)^2+2\mu_+({\cp k}^2/2m+2\Omega)}$, with $\mu_{\pm}=gn(1\mp \eta)/2$. Clearly, in the mean-field collapse regime with $\eta<-1$, the lower spectrum $E_{1{\cp k}}$ becomes purely imaginary near ${\cp k}\sim 0$, signifying the dynamical instability. In addition, we note that under certain condition the two spectra become degenerate, i.e., $E_{1{\cp k}_0}=E_{3{\cp k}_0}$ at: 
\begin{equation}
|{\cp k}_0|=\sqrt{2m\Omega((\eta-2\tilde{\Omega})^{-1}-1)}, \ \ \ {\rm if}\ \ \ 0<\eta-2\tilde{\Omega}<1. \label{condition}
\end{equation}
The according plot is given in Fig.\ref{PT_spectrum} (a). This feature will lead to interesting excitation property when turn on $\gamma$. 

\begin{figure}[t]
\includegraphics[width=8.5cm]{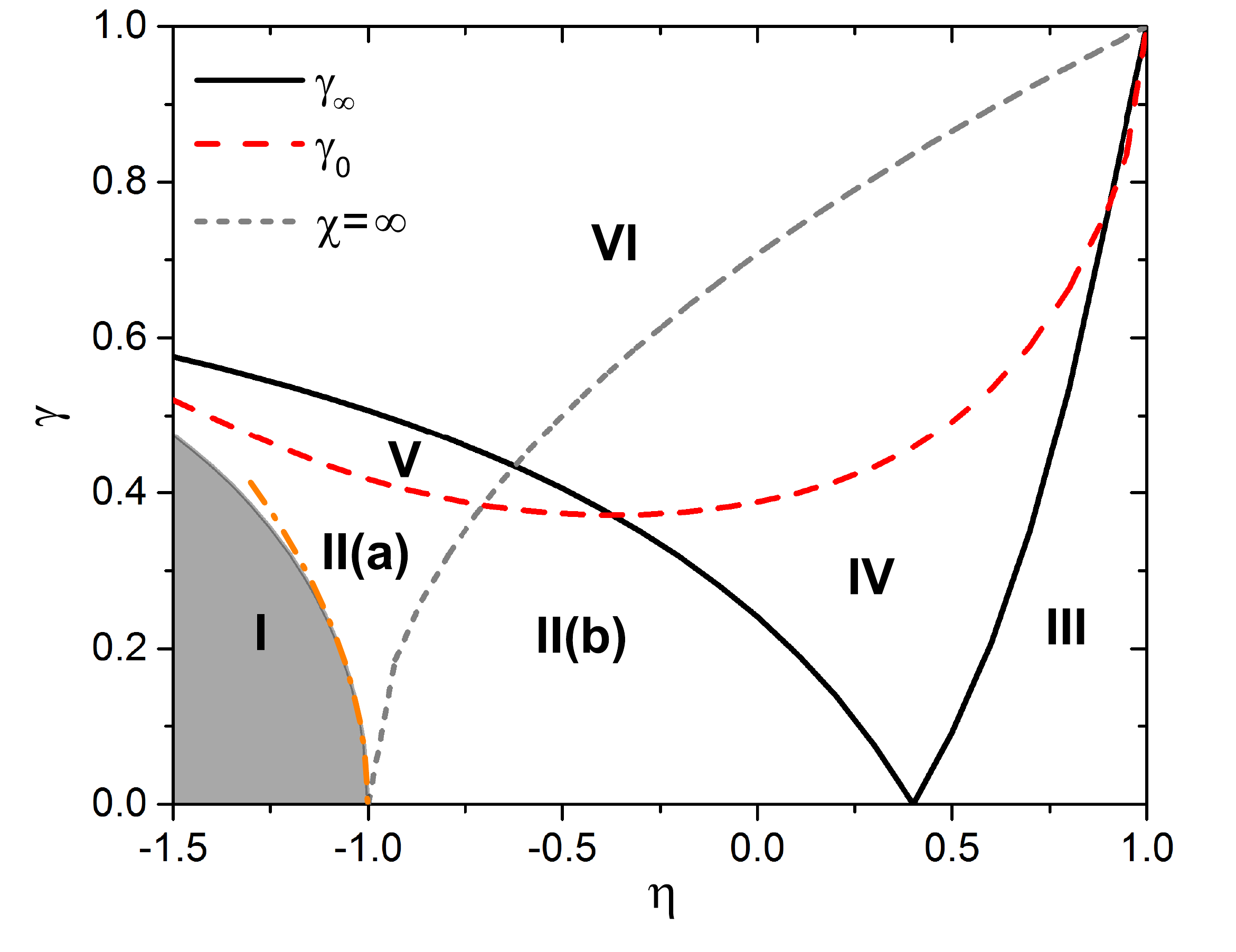}
\caption{(Color online). Diagrams in $(\gamma,\ \eta)$ plane that exhibit different excitation properties. Here $\tilde{\Omega}=0.2$.  'I' marks the region where the excitation spectrum at low-${\cp k}$ is purely imaginary. 'II' is the region where all spectra are real and gapped. The gray dashed line sets the mean-field collapse boundary, which further divides II into II(a) ($\chi<0$) and II(b) ($\chi>0$). $\mathcal{PT}$-breaking transition of Bogoliubov modes occurs in regions III, IV and V, where the spectra become complex either within intermediate  $|{\cp k}|\equiv k$(III), or at large $k$(IV), or low $k$(V). In VI, the complex spectra occur for all ${\cp k}$. These regions are separated by the curves of $\gamma_0$ and $\gamma_{\infty}$, which, respectively, are the values of $\gamma$ when the spectra becomes complex  at $k=0$ and $k\rightarrow\infty$.  } \label{phase_diagram}
\end{figure}

In the presence of non-Hermiticity($\gamma\neq 0$), the inter-branch fluctuations give two important impacts on the Bogoliubov modes, namely, the spontaneous  $\mathcal{PT}$-symmetry breaking and the gapped excitation, as detailed below.

\subsection{Spontaneous  $\mathcal{PT}$-symmetry breaking} 

Although $\mathcal{PT}$-symmetry is preserved by $H, \ H_{\rm BG}$ and the condensate $|\Psi_0\rangle_{R,L}$, it can be spontaneously broken by the Bogoliubov quasi-particles, as manifested by the appearance of complex $E_{i{\cp k}}$. The $\mathcal{PT}$-broken region in ${\cp k}$-space sensitively depends on parameters $\tilde{\Omega},\ \eta$ and $\gamma$. In Fig.\ref{phase_diagram}, we have divided   $(\gamma,\ \eta)$ plane into different regions (I-VI)   according to different $\mathcal{PT}$-breaking properties in the Bogoliubov spectra for a fixed $\tilde{\Omega}=0.2$. The complex spectra occur in regions III-VI. 

\begin{figure}[t]
\includegraphics[width=9cm, height=11cm]{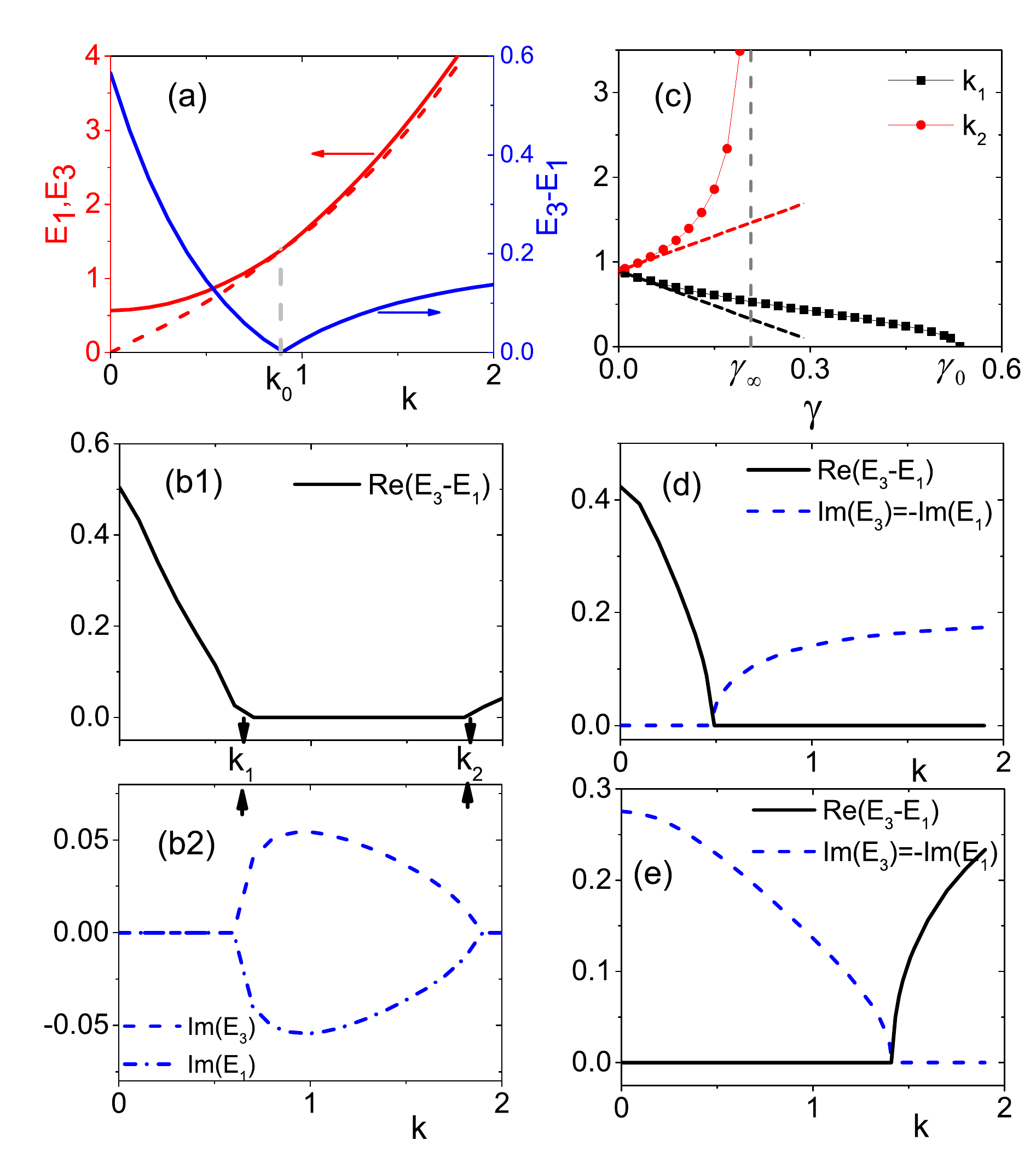}
\caption{(Color online). Spontaneous $\mathcal{PT}$-symmetry breaking in the Bogoliubov spectra. Here $\tilde{\Omega}=0.2$, and $\eta=0.6$ for (a-c). (a) Two real spectra for Hermitian case ($\gamma=0$), which merge at $k\equiv|{\cp k}|=k_0$(see Eq.\ref{condition}). (b1,b2) Real and imaginary parts of the spectra at $\gamma=0.15$ (staying in region III), which shows $\mathcal{PT}$-symmetry breaking for $k\in (k_1,k_2)$. (c) $\mathcal{PT}$-breaking boundaries $k_1$ and $k_2$ as functions of $\gamma$. Dashed lines show linear fits for small $\gamma$ (see Eq.\ref{linear}). $k_1$ touches zero at $\gamma_0$ and $k_2$ goes to $\infty$ at $\gamma_{\infty}$. (d) and (e): Excitation spectra for $\eta=0.4, \gamma=0.3$ (region IV) and for $\eta=-0.8, \gamma=0.45$ (region V), where the $\mathcal{PT}$-breaking occurs, respectively, at high $k$ and low $k$. Here the momentum and energy units are, repectively, $\sqrt{2mgn}$ and $gn$.  } \label{PT_spectrum}
\end{figure}

Let us start from region III  with a small $\gamma$ and $2\tilde{\Omega}<\eta<1$ (satisfying the condition in (\ref{condition})). In this case, a finite $\gamma$ will lead to the $\mathcal{PT}$-breaking of excitation spectra near ${\cp k}_0$. As shown in Fig.\ref{PT_spectrum}(b1,b2), for $\gamma=0.15$, $E_{1{\cp k}}$ and $E_{3{\cp k}}$ are complex and conjugate to each other within a finite window $|{\cp k}|\equiv k\in(k_1,k_2)$. Thus, as increasing $k$ from zero, the $\mathcal{PT}$-symmetry breaks at $k_1$ and then revives at $k_2$. %, where the real parts of $E_{1{\cp k}}$ and $E_{2{\cp k}}$ merge together and their imaginary parts start to develop; this symmetry revives as further increasing $k$ to $k_2$, where the imaginary parts vanish and the spectra regains a real gap in-between. 
The critical boundaries $k_1,\ k_2$, which are determined by the solutions to $b_{\cp k}^2-4c_{\cp k}=0$,  sensitively depend on $\gamma$, see Fig.\ref{PT_spectrum}(c). At small $\gamma$, we find that $k_{1,2}$ deviate from $k_0$ by a small shift $\delta\equiv |k-k_0|$, with
\begin{align}
&\frac{\delta}{k_0}=\gamma \frac{(1-\eta)\sqrt{16\tilde{k}_0^4 +2\tilde{k}_0^2(4+3\eta+10\tilde{\Omega})+(1+\eta) (2\tilde{\Omega}-\eta+1) }}{4\tilde{k}_0^2 (\eta-2\tilde{\Omega})}\nonumber\\
&\ \ \ \ \ \ \ \ \  + o(\gamma^2). \label{linear}
\end{align}  
Here $\tilde{k}_0=k_0/\sqrt{2mgn}$. As shown by the dashed lines in Fig.\ref{PT_spectrum}(c), the dominant linear shifts based on above equation fit well to $k_{1,2}$ in small $\gamma$ limit.  

Continuously increasing $\gamma$, $k_2$ and $k_1$ respectively flow to $\infty$ and $0$ at $\gamma_{\infty}$ and $\gamma_0$. This tells that the spectra at large $k$ become complex if $\gamma>\gamma_{\infty}$,  and the complex spectra extend to $k=0$ if $\gamma>\gamma_{0}$. Numerically, $\gamma_0$ is determined by satisfying $b_{k=0}^2=4c_{k=0}$. To find out $\gamma_{\infty}$ accurately, we expand the function $F_{\cp k}\equiv b_{\cp k}^2-4c_{\cp k}$ at large $k\rightarrow \infty$ and only keep its leading order $\sim k^4$. Then  $\gamma_{\infty}$ is determined by the coefficient of this leading term crossing zero, which gives the equation:
\begin{equation}
\eta^2+\frac{4\gamma_{\infty}^2(\eta-1)}{1-\gamma_{\infty}^2}+4\tilde{\Omega}^2(1-\gamma_{\infty}^2)=\frac{4\tilde{\Omega}(\eta+\gamma_{\infty}^2(\eta-2))}{\sqrt{1-\gamma_{\infty}^2}}.
\end{equation}
We can see that the above equation support a solution $\gamma_{\infty}=0$ at $\eta=2\tilde{\Omega}$. This is also consistent with Eq.(\ref{condition}), which tells that the degenerate point $k_0$ goes to $\infty$ in the Hermitian case if $\eta=2\tilde{\Omega}$.

In Fig.\ref{phase_diagram}, $\gamma_0$ and $\gamma_{\infty}$ are plotted as functions of $\eta$, and accordingly regions III-IV are separated. Specifically, %, IV, V and VI. According to the definitions of $\gamma_0$ and $\gamma_{\infty}$, we can see that the complex spectra, or 
the $\mathcal{PT}$-breaking of Bogoliubov modes occur within a finite $k$-window in  III (with $\gamma<\gamma_0,\gamma_{\infty}$),  at large $k$ in IV ($\gamma_{\infty}<\gamma<\gamma_{0}$), at small $k$ in V ($\gamma_{0}<\gamma<\gamma_{\infty}$), and extend the whole $k$-space in VI($\gamma>\gamma_0,\gamma_{\infty}$). The typical spectra in regions IV and V are given in Fig.\ref{PT_spectrum}(d,e). 
Therefore, the $\mathcal{PT}$-breaking transition takes place twice in III, once in IV and V, and no transition in VI. 
This shows that the $\mathcal{PT}$-symmetry of Bogoliubov modes can be conveniently tuned by $\gamma$ and $\eta$. 

\subsection{Gapped excitation} 

\begin{figure}[t]
\includegraphics[width=8cm]{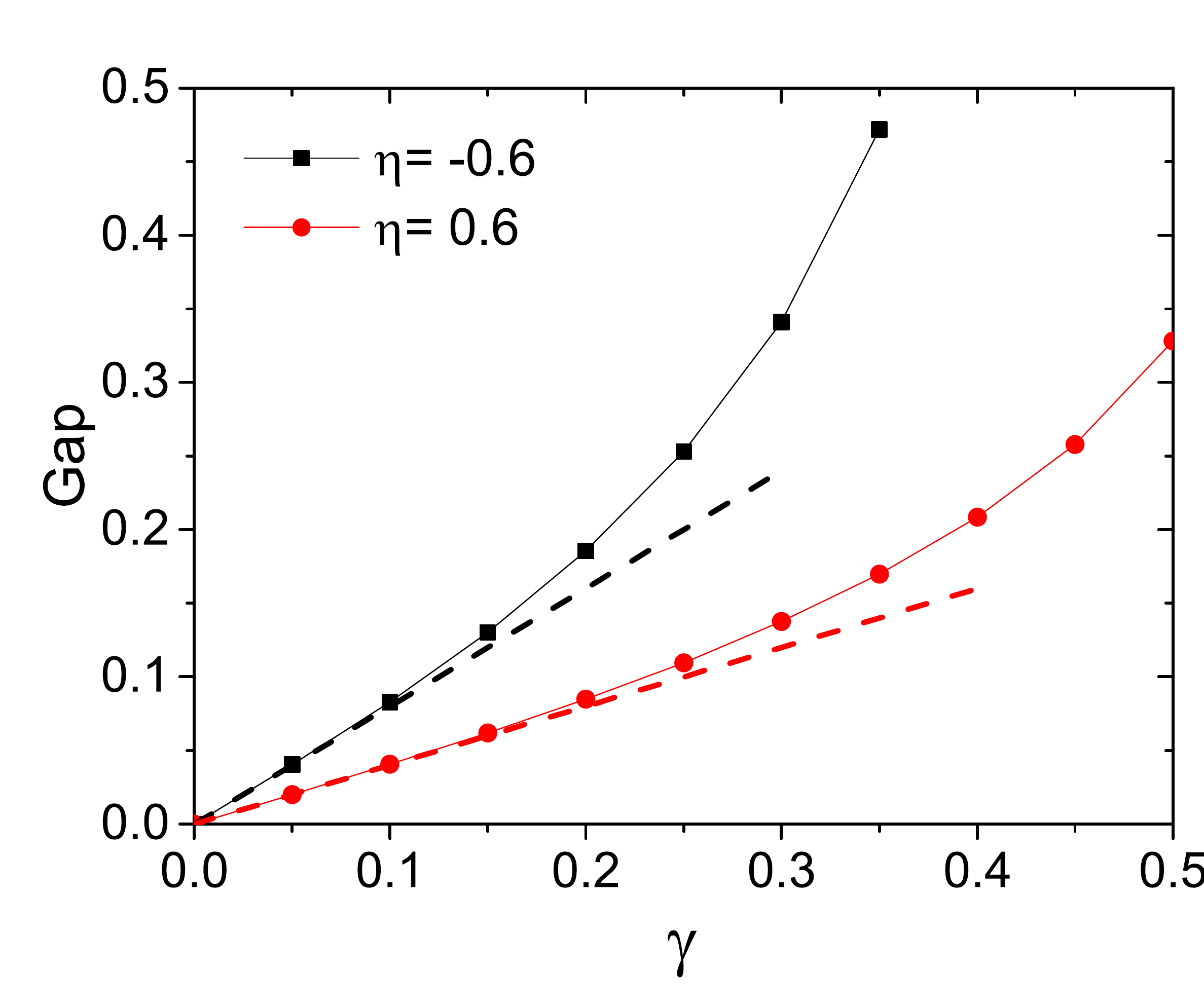}
\caption{(Color online). Excitation gap as a function of $\gamma$ for $\eta=-0.6$ and $0.6$. Dashed lines show the respective function fit according to Eq.\ref{gap}. Here $\tilde{\Omega}=0.2$. The energy unit is $gn$.} \label{fig3}
\end{figure}

In the $\mathcal{PT}$-unbroken region, such as II in Fig.\ref{phase_diagram}, the real Bogoliubov modes are gapped, instead of gapless as in Hermitian case. For $\gamma\ll 1$, we find that the excitation gap scales linearly with $\gamma$:
\begin{equation}
\frac{E_{1{\cp k}=0}}{gn}= \gamma \frac{1-\eta}{2} \sqrt{\frac{1+\eta}{2\tilde{\Omega}}}. \label{gap}
\end{equation}
Such a gapped spectrum is in distinct contrast to the gapless  mode in the Hermition BEC. It is closely related to the  presence of imaginary odd-parity terms in (\ref{Bog}), such as $\Psi^{\dag}_{+; {\cp k}, R}\Psi_{-; {\cp k}, L}$ which directly couple the condensed atoms at $'-'$-branch with higher $'+'$-branch crossing a finite energy gap. Such coupling takes no effect for a $\mathcal{PT}$-symmetric BEC in the mean-field level but plays an important role in its quantum fluctuations. Because such imaginary coupling only exists for $\gamma\neq 0$ and $\eta\neq 1$, the quasi-particle is gapped in the same regime (see (\ref{gap})).    %, as also evidenced in (\ref{gap}). % in (\ref{}) The existence of these asymmetric channels can be attributed to the interplay effect of non-Hermiticity($\gamma\neq 0$), which gives the non-orthogonality of single-particle eigen-modes, and the interaction anisotropy ($\eta\neq 1$). 
In Fig.\ref{fig3}, we extract the energy gap as a function of $\gamma$ for two typical $\eta$, which fit well to (\ref{gap}) in small $\gamma$ regime.

Interestingly, the gapped excitation appears not only in the mean-field stable regime (region II(b)), but can also extend to the collapse regime (II(a)). This is in distinct contrast to the Hermitian case where the low-$k$ spectrum is purely imaginary in the mean-field collapse side. It is to say, the mean-field instability in non-Hermitian system does not necessarily lead to imaginary excitations. In fact, for a given $\eta<-1$, the excitation spectra can turn from purely imaginary to purely real as increasing  $\gamma$ across a critical $\gamma_c$, at which point the spectrum is gapless $E_{1,{\cp k}=0}=0$. In Fig.\ref{phase_diagram}, we mark the $\gamma<\gamma_c$ region ('I') as shaded area, where the low-energy excitation spectra are purely imaginary. 

Numerically, $\gamma_c$ is determined by $c_{{\bf k}=0}=0$, and thus
\begin{equation}
\frac{9\gamma_c^2(1-\eta)^2}{4}=-(1+\eta-2\gamma_c^2)\left[ 2\tilde{\Omega}(1-\gamma_c^2)^{3/2} +(1-\eta)(1+\gamma_c^2) \right].
\end{equation}
We can see that $\gamma_c=0$ when $\eta=-1$, reproducing the mean-field collapse point for the Hermitian case. When $\eta$ slightly deviates from $-1$, we have  
\begin{equation}
\gamma_c=\sqrt{-(1+\eta)\frac{2\tilde{\Omega}+2}{5-4\tilde{\Omega}}},
\end{equation}
which shows that $\gamma_c$ scales as the square root of the deviation, as displayed by the orange dash-dot line in Fig.\ref{phase_diagram}.

\section{$\gamma$-induced droplet}  \label{droplet}

\begin{figure}[t]
\includegraphics[width=8.5cm]{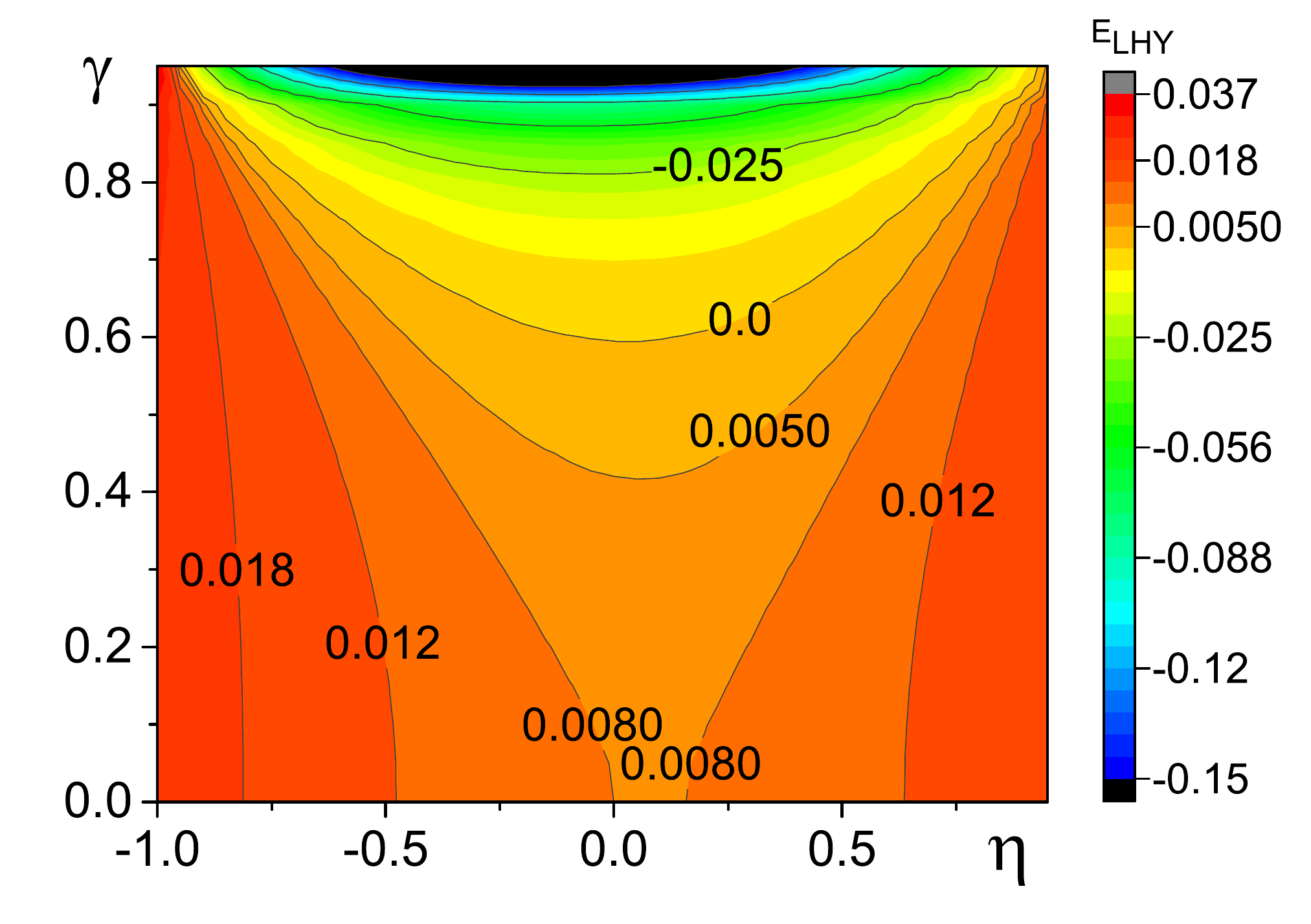}
\caption{(Color online). Contour plot of $f$-function (see Eq.\ref{f}) in the $(\gamma, \eta)$ plane with $\tilde{\Omega}=0.2$. } \label{fig4}
\end{figure}

The fact that the non-Hermiticity $\gamma$ enhances the mean-field collapse (as inferred by Eq.(\ref{stable})) renders the formation of a self-bound droplet after incorporating the LHY correction from quantum fluctuations. In general, Eq.(\ref{E_LHY}) gives ${\cal E}_{\rm LHY}\equiv E_{\rm LHY}/V$ as:
\begin{equation}
{\cal E}_{\rm LHY}=(2m)^{3/2}(gn)^{5/2} f(\gamma, \eta, \tilde{\Omega})  \label{f}
\end{equation}
where $f$ is a dimensionless  functional. In Fig.\ref{fig4}, we show the contour plot of $f$ in $(\gamma, \eta)$ plane given a fixed $\tilde{\Omega}=0.2$. We can see that $f$, or equivalently ${\cal E}_{\rm LHY}$, decreases continuously as $\gamma$ increases and can even turn negative. Fortunately, in region II(a), which is the mean-field collapse regime with real and gapped  spectra, the LHY force is always repulsive. A self-bound droplet state can then be supported in this region with zero pressure, i.e., $\partial(E/N)/\partial n=0$, with $E=E_{\rm mf}+E_{\rm LHY}$. This gives the equilibrium density of the droplet as
\begin{equation}
n_{\rm eq}=\left(\frac{1-2\gamma^2+\eta}{1-\gamma^2}\right)^2\frac{1}{36(2mg)^{3}f^2(\gamma, \eta, \tilde{\Omega})}.
\end{equation}
%One can see that $n_{\rm eq}$ can be conveniently tuned by $\gamma$ and $\eta$. 

At small particle number $N$, the quantum pressure becomes important and drives the droplet to gas transition. To estimate the critical number $N_c$ at the transition, we take the similar strategy as in Ref.\cite{Petrov} and write down the extended Gross-Pitaevskii(GP) equation as
\begin{equation}
i\partial_t \Psi({\bf r})= \left(-\frac{1}{2m} \nabla_{\bf r}^2 + \frac{1-2\gamma^2+\eta}{2(1-\gamma^2)} g |\Psi|^2 + \frac{\partial {\cal E}_{\rm LHY}}{\partial n} \right) \Psi({\bf r}),
\end{equation} 
where $\Psi({\bf r})$ is the wave function of the BEC and the particle number is determined by $N=\int d^3{\bf r}|\Psi({\bf r})|^2$. By rescaling ${\bf r},\ \Psi, \ t$ through
\begin{equation}
{\bf r}=\tilde{\bf r} \xi, \ \ \ \ \Psi=\tilde{\Psi} \sqrt{n_{\rm eq}}, \ \ \ \ \ t=\tilde{t} m\xi^2,  \label{scale}
\end{equation} 
with 
\begin{equation}
\xi=\sqrt{\frac{6(1-\gamma^2)}{mgn_{\rm eq}|1-2\gamma^2+\eta|}},
\end{equation}
we can reduce the GP equation to 
\begin{equation}
i\partial_{\tilde{t}} \tilde{\Psi}(\tilde{\bf r})= \left(-\frac{1}{2} \nabla_{\tilde{{\bf r}}}^2 -3 |\tilde{\Psi}|^2 + \frac{5}{2} |\tilde{\Psi}|^3  \right) \tilde{\Psi}(\tilde{\bf r}). \label{GP}
\end{equation} 
It is found that Eq.\ref{GP} shares the same structure as the reduced GP equation in Hermitian case\cite{Petrov}, which leads to the rescaled critical  number $\tilde{N}_c\equiv\int d^3\tilde{\bf r} |\tilde{\Psi}|^2=18.65$ at the vanishing of droplet solution (droplet-gas transition). Given the scaling relation in (\ref{scale}), we can obtain the critical $N_c=n_{\rm eq} \xi^3 \tilde{N}_c$ in our system as
\begin{equation}
N_c=2\sqrt{2}f \tilde{N}_c \left(\frac{6(1-\gamma^2)}{|1-2\gamma^2+\eta|}\right)^{5/2}.
\end{equation} 
We can see that both $n_{\rm eq}$ and $N_c$ can be conveniently tuned by $\gamma$ and $\eta$. % obtained by simuadopting the single-mode approximation and mapping the system to single-species bosons, which is found to follow the same relation to $n_{\rm eq}$ as in the Hermitian case\cite{Petrov}.

\section {Experimental relevance}  \label{expt}

A $\mathcal{PT}$-symmetric two-species BEC can be realized using two hyperfine states of $^{87}$Rb bosons, $|\uparrow\rangle=|F=1,m_F=1\rangle$ and $|\downarrow\rangle=|F=2,m_F=-1\rangle$. The intra-species scattering lengths are $a_{\uparrow\uparrow}=95a_B,\ a_{\downarrow\downarrow}=100a_B$, which have very small relative asymmetry  $|a_{\uparrow\uparrow}-a_{\downarrow\downarrow}|/(a_{\uparrow\uparrow}+a_{\downarrow\downarrow})\sim 2.5\%$. Such a small asymmety is expected to take little effect as long as it is much smaller than $\tilde{\Omega},\ \gamma$. The inter-species coupling is highly tunable via Feshbach resonance around $B_0=9.1$G\cite{FR_Rb1,FR_Rb2,FR_Rb3}. 

For the $\mathcal{PT}$-symmetric potential (\ref{H_PT}), the  $\sigma_x$ term can be implemented %by $|\uparrow\rangle$ and $|\downarrow\rangle$ can be coupled 
through the two-photon microwave and rf transition\cite{Rb_rf}, and $i\sigma_z$ can be realized using the laser-induced state-selective dissipation up to a constant loss term $i\Omega\gamma$\cite{Luo, Jo, Gadway}. 
 For realistic atomic system with such constant loss, the overall number of the system decays with time. However, the physics governed by the effective non-Hermitian Hamiltonian can still be probed under the post-selection scheme, as have been successfully explored in the non-interacting atomic gases\cite{Luo, Jo}. The validity of the effective non-Hermitian Hamiltonian requires a short-time dynamics within timescale $t\ll 1/(\Omega\gamma)$, where the impact of quantum jump  can be neglected. As here we consider the weak coupling regime with $na^3\ll1$, which is a natural extension of and can be smoothly connected to the non-interacting regime, we do not expect the validity of the effective non-Hermitian Hamiltonian would alter too much. Moreover, it should be noted that the existing experiments on quantum droplet have exactly made use of the atom loss to observe the droplet-gas transition\cite{Tarruell_1,Tarruell_2,Inguscio, Modugno_2, Wang}. We thus expect that the $\gamma$-induced droplet can be directly probed in realistic experiments. 

The property of excitation spectrum %, including the real gap and the spontaneous $\mathcal{PT}$-symmetry breaking, 
can be explored by the Bragg spectroscopy as implemented previously in various cold atoms systems\cite{excitation_expt1,excitation_expt2, excitation_expt3,excitation_expt4}. 
Since such spectroscopy detects the linear response of the system to external perturbations, we expect it can directly probe the excitation  spectrum of non-Hermitian system as predicted in this work. 
Our results, which are directly relevant to atomic gases confined in a uniform trap\cite{box_trap1,box_trap2, box_trap3, box_trap4, box_trap5}, can also be utilized for the trapped system under  local density approximation, as successfully implemented in previous experiments \cite{excitation_expt1,excitation_expt2,excitation_expt4}. %Finally, we expect the spectroscopy measurement can directly pro bi-orthogonal framework our results are directly relevant to atomic gases in a uniform trap, as has been realized in recent years\cite{box_trap1,box_trap2, box_trap3, box_trap4, box_trap5}. 

\section{Summary and discussion}  \label{summary}

In summary, we have revealed the ground state and excitation properties of a $\mathcal{PT}$-symmetric BEC,  including the spontaneous $\mathcal{PT}$-breaking and gapped spectrum for Bogoliubov quasi-particles, and the enhanced mean-field collapse and the facilitated  droplet formation. %These properties are in direct contrast to the Hermitian case, thereby demonstrating the significant interplay of interaction and non-Hermiticity in BEC system. Moreover, we emphasize that these properties 
These results show that the quantum fluctuations on top of a $\mathcal{PT}$-symmetric BEC can lead to important and visible collective phenomena  even far from the single-particle EPs, thus demonstrating the significant interplay of interaction and non-Hermiticity in bosonic system.

%In particular, the intriguing excitation properties are closely related to the presence of odd-parity coupling channels, which can be traced back to the fundamental character of non-Hermitian systems, i.e., the non-orthogonality of eigenstates. 
Finally, we point out %that the intriguing excitation properties revealed in this work BEC, as shown in Figs.\ref{phase_diagram}-\ref{fig3}, robustly exist in a wide parameter regime ($\gamma<1,\eta<1$). This manifests the dramatic effect of quantum fluctuations to the PT-symmetric BEC  even far from the single-particle EP. 
 that the intriguing excitation properties revealed in this work can be traced back to the fundamental character of non-Hermitian systems, i.e., the non-orthogonality of eigenstates. Such non-orthogonality character covers both the single-particle states and the elementary quasi-particles. This is why the $\mathcal{PT}$-symmetry breaking can also occur in the latter. % activate the inter-branch coupling channels and lead to intriguing excitation property such as the real gap and the spontaneous $\mathcal{PT}$-symmetry breaking in the excitation spectrum.  Since the e 
We thus expect the phenomena revealed here are  not limited to the specific $\mathcal{PT}$-potential considered in this work, but applicable to a broad class of non-Hermitian systems with $\mathcal{PT}$-symmetry. Indeed, a recent study has pointed out the spontaneous $\mathcal{PT}$-breaking of elementary excitations on top of a fermion superfluid\cite{Yi}. These $\mathcal{PT}$-breaking phenomena are generally associated with the collective many-body EP and may lead to giant fluctuation effect\cite{Littlewood,Littlewood2}. In future, it is worth to explore the impact of collective EPs in the quantum and thermal depletions, as well as the property of BEC in other parameter regime ($\gamma,\eta,\tilde{\Omega}$)  beyond the scope of this work. 

\bigskip

{\bf Acknowledgement.} We thank Dajun Wang for helpful discussion on experimental realization of the system. The work is supported by the National Key Research and Development Program of China (2018YFA0307600), the National Natural Science Foundation of China (No.12074419), and the Strategic Priority Research Program of Chinese Academy of Sciences (No. XDB33000000).

\bigskip

\appendix

\section{Heisenberg equation for non-Hermitian system} \label{Heisenberg}

We first derive the Heisenberg equation for non-Hermitian system under the bi-orthogonal basis. 
Given the definition of right and left states, at time $t$ they evolve as 
\begin{align}
&|\phi_R(t)\rangle=e^{-iHt}|\phi_R(0)\rangle,\\
&|\phi_L(t)\rangle=e^{-iH^\dagger t}|\phi_L(0)\rangle,
\end{align}
here $|\phi_{R}(0)\rangle$ ($|\phi_{L}(0)\rangle$) is the initial right (left) state at $t=0$. Define the time-dependent expectation value of operator $\hat{A}$ as
\begin{equation}
\langle \hat{A} \rangle_t \equiv \langle \phi_L(t) |\hat{A}|\phi_R(t)\rangle,
\end{equation}
we then have
 \begin{equation}
\langle \hat{A} \rangle_t= \langle \phi_L(0) |e^{iHt}\hat{A} e^{-iHt}|\phi_R(0)\rangle,
\end{equation}
and thus the Heisenberg equation can be written as
\begin{equation}
i \frac{\partial}{\partial t} \langle \hat{A} \rangle_t = \langle [\hat{A}, H] \rangle_t. \label{EoM}
\end{equation}
We can see that the form of Heisenberg equation (\ref{EoM})  is identical to the Hermitian case. Nevertheless, it has a remarkable consequence for the non-Hermitian case, i.e., $\langle \hat{A}^{\dag} \rangle_t \neq \langle \hat{A} \rangle^{*}_t $, which is very different from the Hermitian case. Similar relation for the time-dependent non-Hermitian operators has been given in Ref.\cite{Zhou2}.

\end{document}